\documentstyle[12pt]{article}

\begin{document}


        \newcommand{\be}{\begin{equation}}
        \newcommand{\ee}{\end{equation}}
        \newcommand{\ba}{\begin{eqnarray}}
        \newcommand{\ea}{\end{eqnarray}}
        \newcommand{\ban}{\begin{eqnarray*}}
        \newcommand{\ean}{\end{eqnarray*}}
        \newcommand{\barr}{\begin{array}}
        \newcommand{\earr}{\end{array}}

        \renewcommand{\H}{{\cal H}}
        \newcommand{\K}{{\cal K}}

        \newcommand{\et}{\hspace{-0.08in}{\bf .}\hspace{0.1in}}
        \newcommand{\tensor}{\otimes}
        \newcommand{\maps}{\colon}
        \newcommand{\vol}{{\rm vol}}
        \newcommand{\iso}{\cong}
        \newcommand{\tr}{{\rm tr}}
        \newcommand{\sign}{{\rm sign}}
        \newcommand{\M}{{\it M }}
        \newcommand{\id}{{\rm id}}
        \newcommand{\Diff}{{\rm Diff}}
        \newcommand{\End}{{\rm End}}
        \newcommand{\V}{{\rm Vect}}
        \newcommand{\Vect}{{\rm Vect}}
        \newcommand{\Aut}{{\rm Aut}}
        \newcommand{\Ad}{{\rm Ad}}
        \newcommand{\Maps}{{\rm Maps}}
        \newcommand{\hf}{{1\over 2}}
        \renewcommand{\j}{\jmath}

        \textwidth 6in
        \textheight 8.5in
        \evensidemargin .25in
        \oddsidemargin .25in
        \topmargin .25in
        \headsep 0in
        \headheight 0in
        \footskip .5in

       \parskip 1.75\parskip plus 3pt minus 1pt
        \pagestyle{plain}
        \pagenumbering{arabic}

\def\nn{\nonumber}
\def\bs{\bigskip}
\def\no{\noindent}
\def\hb{\hfill\break}
\def\qq{\qquad}
\def\bl{\bigl}
\def\br{\bigr}
\def\ra{\rightarrow}
\def\Ra{\Rightarrow}
\def\lra{\leftrightarrow}

\def\k{\kappa}
\def\r{\rho}
\def\a{\alpha}
\def\b{\beta}
\def\B{\Beta}
\def\g{\gamma}
\def\G{\Gamma}
\def\d{\delta}
\def\D{\Delta}
\def\e{\epsilon}
\def\c{\chi}
\def\th{\theta}
\def\Th{\Theta}
\def\m{\mu}
\def\n{\nu}
\def\na{\nabla}
\def\om{\omega}
\def\Om{\Omega}
\def\l{\lambda}
\def\L{\Lambda}
\def\s{\sigma}
\def\S{\Sigma}
\def\p{\phi}
\def\P{\Phi}
\def\varp{\varphi}
\def\pa{\partial}
\def\pap{\partial_+}
\def\pam{\partial_-}
\def\papm{\partial_{\pm}}
\def\pamp{\partial_{\mp}}
\def\pat{\partial_{\tau}}
\def\J{{\cal J}}
\def\RN{{Reissner-Nordstr\"{o}m }}

\def\frac#1#2{{#1\over#2}}
\def\pd#1#2{{\frac{\partial#1}{\partial#2}}}

        \newcommand{\Z}{{\bf Z}}
        \newcommand{\Q}{{\bf Q}}
        \renewcommand{\div}{\nabla\cdot}
        \newcommand{\curl}{\nabla\times}
        \newcommand{\grad}{\nabla}
        \newtheorem{ex}{Example}
        \newcommand{\bex}{\begin{ex}\et}
        \newcommand{\eex}{\end{ex}}
        \def\dia{\diamond}
        \def\we{\wedge}
	\def\RN{{Reissner-Nordstr\"{o}m }}
        \newcommand{\ed}{\end{document}}

        \def\sqr#1#2{{\vcenter{\vbox{\hrule height.#2pt
                \hbox{\vrule width.#2pt height#1pt \kern#1pt
                   \vrule width.#2pt}
                \hrule height.#2pt}}}}

        \def\square{\mathchoice\sqr34\sqr34\sqr{2.1}3\sqr{1.5}3}


\newcommand{\A}{{\Bbb {A}}}
\newcommand{\Op}{{\Bbb {O}}}
\newcommand{\lap}{{{}_{{}_{{}_{\sim}}} \!\!\!\!N}}
\newcommand{\R}{{\rm {I\!R}}}
\def\E{{\widetilde{E}}}
\def\doble{{{\lim_{{d\to 0}\atop{\e\to 0}}}}}
\def\N{{\;\lap (x)\;}}
\def\C{{\cal C}_Q}
\def\bh{{black holes }}
\def\rat{{r^2\over{l^2}}}
\def\ram{{{{r_+}^2}\over{l^2}}}
\def\sc{{Schwarzschild }}
\def\rc{{r_+^c}}

\newcounter{figg}
\newcommand{\figura}[2]{\begin{center}
\mbox{\epsfig{file={#1},height={#2cm}}}\end{center}
\stepcounter{figg}}


%
%

        \begin{center}
	{\Large Critical Behavior of \\
		Dimensionally Continued Black Holes \\ }
        \vspace{0.5cm}
	Javier P.\ Muniain$\,{}^{\dagger}$ and Dardo
	P\'{\i}riz$\,{}^{\ddagger}$  \\
	{\small{\it Department of Physics  \\
	University of California, Riverside \\
 	California 92521-0413, U.\ S.\ A.  \\ }}
	\vspace{0.4cm}
	\end{center}

\begin{abstract}
{\footnotesize
The critical behavior of black holes in even and odd dimensional spacetimes
is studied based on Ba\~nados-Teitelboim-Zanelli (BTZ) dimensionally continued
black holes. In even dimensions it is found that asymptotically flat and
anti de-Sitter Reissner-Nordstr\"om black holes present up to two second
order phase transitions. The case of asymptotically
anti-de-Sitter Schwarzschild black holes present only one critical transition
and a minimum of temperature, which occurs at the transition. Finally, it
is shown that phase transitions are absent in odd dimensions.}
\end{abstract}

\vspace{3cm}
{\footnotesize ${}^{\dagger}$muniain@phyun0.ucr.edu \qq
${}^{\ddagger}$dardo@phyun0.ucr.edu}
\vfill\eject

\begin{center}
\section*{I. Introduction}
\end{center}

The possibility of critical behavior and scaling of classical
objects such as black holes in general relativity is an
interesting and open question.
Scaling behavior was discovered by Choptuik \cite{cho}
in connection with the numerical study of gravitational
collapse of massless scalar fields.
In that paper, a universal behavior of the black-hole mass
described by a critical exponent $\b\sim 0.37$ independent
of the initial shape of the collapsing scalar field was found.
Since then, critical behavior and scaling in other collapsing
systems have been reported \cite{as}.

To study the thermodynamics of black holes, and in particular their heat
capacity and critical behavior, it is assumed that there is an existing
analogy between the laws of thermodynamics and the laws that govern
black hole mechanics derived from general relativity. This
was first established by Bardeen, Carter and Hawking \cite{BCH}.
To guarantee this analogy one needs to make the formal
replacements $E\to M$, $T\to c\,\kappa$ and
$S\to A/8\pi c$, where $A$ is the area, $\kappa$ de surface gravity and
$c$ is a constant \cite{beken}. With these
substitutions, the four laws of black hole thermodynamics can be enunciated
and the study of critical behavior seems to be a plausible natural
extension of these ideas. Two early contributions to the study
of critical  behavior in gravitational  systems are those of
Davies \cite{dav}, and Hut \cite{hut}, who
discussed phase transitions in Kerr-Newman and \RN black-holes
in four spacetime dimensions.
In the same direction, Lousto \cite{Lousto}
has argued in favor of the validity of the scaling laws in gravitational
systems. He has calculated the critical
exponents of black holes in four dimensions and has shown the validity
of the scaling laws in those transitions previously found by Davies \cite{dav}.
However, the relationship between the results found in \cite{cho,as} and
\cite{Lousto} are yet to be understood.

Ba\~nados, Teitelboim and Zanelli  have recently reported Schwarzschild and \RN
anti-de-Sitter black hole solutions for even and odd
dimensional spacetime as a particular dimensional continuation
of general relativity with non-vanishing cosmological constant $\L$
\cite{BTZ}. By a suitable choice of coefficients in the Lovelock action
they obtain a unique solution for the metric with dressed
singularity, although only for positive masses. The entropy
becomes a monotonically increasing function
of $r_+$, and therefore the second law of thermodynamics for black holes
remains valid.

\bs
It is our purpose to analytically study the scaling behavior
in gravitational systems and provide further results to compare with
numerical studies in this subject. In this sense, Ba\~nados-Teitelboim-Zanelli
(BTZ) black-holes seem to be an interesting and relatively accesible
arena where to test these ideas. Also the fact that these objects
are defined in general
spacetime dimensions, seems to be a distinctive feature that might help
clarify whether the {\it universality hypothesis} of scaling behavior
is true for  gravitational systems; that is, whether the critical exponents
depend only on the dimensionality of the system, on the dimensionality
of some order parameter and on the range of the gravitational force.
In addition to this, results on the thermodynamics and critical
behavior of \RN anti-de-Sitter black-hole solutions scarce in the
literature.

\bs
In this paper we study the occurrence of phase transitions
in dimensionally continued BTZ black holes.
We review some of the results found in \cite{BTZ},
and particularize them for the different cases where the charge
$Q$ and $\L$ are zero or non-vanishing.
In our study, as in the BTZ paper, we only consider non-rotating ($J=0$)
black holes. The scaling behavior associated to these transitions needs
further study, and we shall report on that elsewhere \cite{mp}.
In section II we briefly introduce the
Lovelock action and the particular choice of coefficients from
where the Schwarzschild and \RN anti-de-Sitter black holes are derived.
In sections III and IV we study the critical behavior of these
\bh in even and odd dimensions respectively, by evaluating the full
thermal capacity at constant $Q$. We find that phase transitions are
possible in even dimensions, except for the case of \sc \bh with zero
cosmological constant. We also obtain that odd dimensional scenarios
do not present transitions. Here, we also study possible discontinuities
in the derivatives of the thermal capacity to assure that there are no phase
transitions of any odd order. Section V is dedicated to give our conclusions.

\begin{center}
\section*{II. Dimensionally continued black holes}
\end{center}

The Lovelock action \cite{love} in $D$ dimensions, which is made by the
sum of the dimensionally continued Euler characteristics of dimensions less
than $D$ \cite{teitel2}, is considered to be the most
general extension of Einstein's gravity
that keeps the field equations for the metric to second order
\footnote{Other
models have been considered in the literature, which include string
theory based black hole solutions and 2-dimensional
black holes with a dilaton field \cite{dilaton}, but we will not go
into those approaches here.}.

The action is written as
\be
S=\k\sum_{p=0}^n \a_p S_p,  \label{action}
\ee
where
\be S_p = \int \e_{a_1\cdots a_D}\,R^{a_1 a_2}\we\cdots\we R^{a_{2p-1}a_{2p}}
\we e^{a_{2p+1}}\we\cdots\we e^{a_D}  \label{action2}.
\ee
Here $R^a_b= d\om^a_b + \om^a_c \om^c_b$ is the curvature 2-form, $\om^a_b$
is the spin connection, and $e_a$ is the local frame one-form.
The action is a local Lorentz invariant $D$-form and is made of $e_a$,
$\om^a_b$ and their exterior derivatives.
However, these conditions do not restrict the values of the
$\a_p$ coefficients. To obtain $\a_p$ Ba\~nados, Teitelboim and Zanelli
\cite{BTZ}
consider the embeddig of the Lorentz group $SO(D-1,1)$ into the anti-de
Sitter group $SO(D-1,2)$. We consider the choice of coefficients $\a_p$
that appears in \cite{BTZ}
\be
\a_p = \left\{\begin{array}{ll}
	  {n\choose p}l^{-D+2p} &\mbox{if $D=2n$}   \\
	  {1\over{D-2p}}{{n-1}\choose p}l^{-D+2p} &\mbox{if $D=2n-1$},
\label{coeff}  \end{array} \right.
\ee
where $l$ is a length related to the cosmological constant by $l=-a^2/\L$
($a>0$). These coefficients are constants
with dimensions $[{\rm mass}]^{D-2p}$, $\k$
has units of action (dimensionless, if $\hbar=1$) and $a_i={0,\ldots, D-1}$.

The authors of \cite{BTZ} restricted their analysis to the cases of
$D=4k$ and $D=4k-1$
($k\in{\bf Z}$) in order to avoid naked singularities with positive mass in the
BTZ model. However, black holes with regular horizons exist in the remaining
dimensions provided one takes only one of the two possible branches of real
solutions, namely that corresponding to the positive root of $\left(2M/r\right)
^{1/(n-1)}$ for $D=2n$, or the positive root of $(M+1)^{1/(n-1)}$
for $D=2n-1$, and similarly for the charged solutions.

\begin{center}
\section*{III. Even dimensional black holes}
\end{center}

In even dimensions the action (\ref{action}) is of the
form
\be
{\cal L}_{\rm even} = \k (R^{a_1 a_2}+ l^{-2} e^{a_1}\we e^{a_2})\we\cdots\we
(R^{a_{D-1}a_D}+ l^{-2} e^{a_{D-1}}\we e^{a_D})\e_{a_1\cdots a_D}, \label{acev}
\ee
and the equations of motion are given by
\be
(R^{a_1 a_2}+ l^{-2} e^{a_1}\we e^{a_2})\we\cdots\we
(R^{a_{D-3}a_{D-2}}+ l^{-2} e^{a_{D-3}}\we e^{a_{D-2}})\we e^{a_D}
\e_{a_1\cdots a_{D-1}}=0.
\ee
The factorized form of these equations,
due to the particular choice of the coefficients,
leads to a much simpler study of the physical
properties of its solutions \cite{BTZ}.

To study spherically symmetric solutions of \bh we start
from the metric
\be	ds^2 = - g^2(r) dt^2 + g^{-2}(r) dr^2 + r^2 d \Om^2,
\label{interval} 	\ee
where the coefficient $g^2(r)$ can be expressed as a function
of $r$, $l$, $M$, and $Q$ as
\be 	g^2(r) =  1 + \rat - [{2M\over r} - {Q^2\over{(D-3)\,
r^{D-2}}}]^{2\over{D-2}}.  \label{metric}	\ee

Spherically symmetric solutions for the Einstein-Lovelock equations
have been studied in the literature by several authors \cite{wheeler};
we consider here only static and spherically symmetric metrics.

The solution of $g^2(r) = 0$ gives us the value of the event horizon
$r=r_+$
\be
{Q^2\over{2M(D-3)}} + {r_+^{D-2}\over{2M}}(1+ \ram)^{(D-2)/2} = r_+^{D-3},
\label{constraint} \ee
which is the constraint relation we will use in following expressions to
write the relevant thermodynamical quantities.
The above equation has zero, one or two real solutions, depending
on the values of $M$, $Q$ and $l$. We will illustrate in some
detail how the roots can be obtained with a reasoning that will be
used in the rest of the paper.

The LHS in (\ref{constraint}) is a polynomial in $r_+$
of degree $2D-4$ with strictly positive coefficients. The RHS is a polynomial
of order $D-3$, with positive coefficient also.
Both sides are monotonically  increasing functions of $r_+$.
The LHS is dominant for small (provided $Q^2/M\ne0$) and large values
of $r_+$.
The RHS may dominate in the intermediate region, depending on the values
of $Q$ and $M$. Then, there will be two, one or zero solutions to
(\ref{constraint}).
For very large values of $Q^2/M$ the LHS is always dominant, and there is
no root, which means that we have a naked singularity. If we decrease the
ratio $Q^2/M$, it will reach a
value for which the LHS and RHS curves will be tangent to each other at
one point. In that case we have only one root and therefore one event horizon.
Below that value of $Q^2/M$ the two curves intersect in two points, and
thus two horizons arise. We take the greater root of the two as the
black hole horizon.
We will use this kind of reasoning later on to get a feeling of how
the system behaves in the general case where explicit solutions
will not be available.

For $l\to\infty$, that is
$\L=0$, and $Q=0$, equation (\ref{constraint}) gives us $r_+ = 2M$
for the case of
\sc black holes, whereas the finite cosmological constant situation,
gives $r_+ < 2M$. Therefore the largest \sc horizon happens
for the $\L=0$ scenario.

 From the standard expression for the temperature
\be	T = {1\over {4 \pi}} \left( {d g^2(r)\over{d r}}\right)_{r=r_+}
\label{def}	\ee
and relation (\ref{constraint}), we find
\be	T = {{r_+^{1-D}(1+\ram)^{(4-D)/2}}\over {2\pi(D-2)}}
\left[ r_+^{D-2}[1+(D-1)\ram](1+\ram)^{(D-4)/2} - Q^2 \right].
\label{temp}	\ee

In the asymptotically anti-de-Sitter \sc case ($Q=0$ and negative $\L$),
it is easy to see that the above expression for the temperature reduces to
the result obtained in \cite{BTZ}
\be   	T_{Q=0} = {1+(D-1) (r_+/l)^2 \over {{2\pi} (D-2) r_+}}.
\label{tempq0}   \ee
In the asymptotically flat \RN solution in arbitrary spacetime dimension
($\L=0$ and $Q\ne 0$), the expression for the temperature reduces to
\be	T_{\L=0}= {1\over {2\pi (D-2)}} \lbrack {1\over{r_+}} -
{Q^2\over{r_+^{D-1}}} \rbrack	\label{temp_inf}	\ee
while for the asymptotically flat \sc case is
\be	T_{\L=Q=0}= {1\over{2\pi (D-2) r_+}} = {1\over{4\pi (D-2) M}}.
\label{temp_0}	  \ee
For $D=4$ these results reproduce the standard relations found in
the literature.

The expression for the entropy of the black hole in even dimensions can be
computed in  closed form, obtaining \cite{BTZ}
\be	S(r_+) = \pi l^2\, [ (1+\ram)^{(D-2)/2} - 1 ],  \label{entropy}	\ee
which for the $l\to\infty$ case reduces to
\be	S_{\L=0}(r_+) = {\pi\over 2} (D-2) r_+^2.	\ee

\bs
\subsection*{Phase transitions}

\bs
For the study of phase transitions we need to assume that the system is
held in equilibrium at some temperature $T$ with a surrounding heat bath.
In $D=4$ and $\L=0$, this condition was proved to be true only for
supermassive black holes ($M\ge10^5M_\odot$)\cite{gib}.
Above this limit, there is not enough energy for the emission of non-zero
rest mass particles and the discharge of the black hole due
to Hawking evaporation is negligible. Hence, the assumption of reversible
transfer of energy at constant charge will be true.

We will consider a small reversible transfer of energy between the hole
and its environment in such a way that the electric charge $Q$
remains unchanged. The heat capacity we calculate
is related to this transfer of energy.

By using expressions (\ref{temp}) and (\ref{entropy}) we obtain
\be 	\C = T\left.{{\pa S}\over {\pa T}}\right|_Q = {T\over\D}\,
[2 \pi^2 (D-2)^2 (1+\ram)^{D-3} r_+^{D+1}],  \label{heat}
\ee
where
\be
\D = [{D-1\over{l^2}}r_+^D - r_+^{D-2}](1+\ram)^{(D-2)/2} +
Q^2 [{(2D-5)\over{l^2}} r_+^2 + D-1] 	\label{den}
\ee
To study the critical behavior in these black holes, we
look for solutions of $r_+ = \rc$ that make the denominator $\D$
vanish in the above expression. We shall divide our study in different cases:

\bs

\no {\bf (1)} Asymptotically flat \sc solution in $D$ dimensions.
This case does not present transitions, since there are no values of $r_+$
that make $\D= 0$.
Taking the limit $l\to\infty$ in eq.(\ref{heat}) we obtain
$\C=-\pi (D-2) r_+^2$, which is always negative for any value of $D$.
In $D=4$, $r_+=2M = 1/(4\pi T)$ leading to $\C=-M/T$. The negative
heat capacity implies that a slight drop in black hole temperature
will cause a further drop as energy is absorved. The process will
continue indefinitely, with the black hole feeding on the surrounding
heat bath. The fact that the temperature of \sc \bh increases as they radiate
energy \cite{dav,hut} is also realized for BTZ black holes, independently
of the dimensionality $D$.

\bs

\no {\bf (2)} Asymptotically flat \RN solutions.
This case corresponds physically to retaining only the highest order
dimensionally continued Euler density in the action (\ref{acev}),
for example, the Einstein-Hilbert term for $D=4$. Making $\D= 0$ in
(\ref{den}) we find
\be	\rc = [Q^{c2} (D-1)]^{1/(D-2)}.	\label{rc}	\ee
As the value of $D$ goes to infinity, it is easy to see
that $\rc\to 1$ independent of the charge of the black hole.
For any dimension equal or greater than four we
find that the value of the critical event horizon becomes smaller
as the dimension of spacetime increases. However, this critical horizon
will never become zero.
Substituting (\ref{rc}) into Eq.(\ref{temp_inf}) we find the value for the
critical temperature to be
\be	 T^c = {1\over{2\pi(D-1) \rc}}.             \ee
 From Eq.(\ref{constraint}) we have
\be      M^c = {\rc\over 2} [1+{1\over{(D-1)(D-3)}}],   \ee
so we can write the critical temperature in terms of
the critical mass as
\be	T^c = {1\over{4\pi M^c}}[{1\over{D-1}} + {1\over{(D-1)^2 (D-3)}}].  \ee
The entropy is
\be S^c = {\pi\over2}(D-2)[Q^{c2}(D-1)]^{2/(D-2)}. \label{entro}	\ee

For the case of $D=4$ black holes, the critical values reduce to
$\rc =Q^c\sqrt{3}$, $T^c = 1/(6\pi\rc) = 1/(9\pi M^c)$ (since $M^c = 2\rc/3$),
and $S^c=\pi \rc^2= 3\pi Q^{c2}$.
These results coincide with those of \cite{Lousto}
where
\be	T^c = \big( 2 \pi M [3+\sqrt{3-q_J}]\big)^{-1}	\ee
and $q_J$ satisfies the following constraint
\[	j_J^2 + 6 j_J + 4 q_J = 3.	\]
Since we are only concerned with the case of non-rotating black holes,
$J=0$ and then $q=3/4$. Hence, one finds $T^c =(9\pi M^c)^{-1}$.

We should note here that $T$, $S$ and $M$ all remain finite through the
transition. Since
\be \D= Q^2(D-1)-r_+^{D-2},	\ee
the heat capacity $C_Q$ presents two branches, going from positive to
negative values through an
infinite discontinuity which we can classify as of second order.

\bs
As we mentioned before, we need to check whether these values lie in the
thermodynamical region, or if superradiant discharge modes become
important. Following Gibbons and Carter \cite{gib}, for $D=4$ we see
that $Q^c/M^c=\sqrt{3}/2\gg m_e/e\approx10^{-21}$ and therefore the
critical point lies within the region for which emission of charged
particles is energetically favored. However,
the condition for the superradiant modes to be neglegible,
$Q^c/M^{c2}\le m_e^2/e$ implies $M^c\ge 10^{48}\approx 10^5 M_\odot$.
This imposes a lower limit for the value of the black hole mass for
spontaneous discharge through superradiant modes to be neglegible
and the previous expressions to give a valid critical transition.
Then for $D=4$ it is possible to have critical transitions in
\RN \bh provided $M^c\ge10^5M_\odot$. We expect that the situation
would be similar for the more general BTZ black holes with $\L\neq 0$
and general $D$. However, it is still necessary to show that the Coulomb
barrier arguments presented in \cite{gib} will still hold for any $D$.

\bs
In the study of the temperature and mass as functions of the event horizon,
the maximum value of the temperature
is reached at the critical point $T^c$ and the minimum
happens at $T=0$, which corresponds to $r_+ = Q^{2/D-2}$.
This value of the horizon takes place for $Q^{2/(D-2)} = 2M (D-3)/(D-2)$,
which is the limiting case of a degenerate horizon. If $Q$ exceeds
this value there is a naked singularity.

In (\ref{entro}) we encounter a limiting value for the zero-point
entropy
\be
S^{(0)} = {\pi\over 2} (D-2) Q^{4/(D-2)},
\ee
or $S_{D=4}^{(0)} = \pi Q^2$. We thus expect the ground state of these
BTZ black holes to be degenerate.

As the black hole horizon goes over $\rc$
given by eq.(\ref{rc}), the black hole temperature decreases, reaching
the zero value for infinite horizon. The mass behaves
as a monotonically increasing function of $r_+$ for those values
that make the temperature positive. The region where the mass is monotonically
decreasing corresponds to negative values of $T$, which lacks of
physical meaning. The heat capacity as a function of the temperature presents
two branches, depending on the value of the event horizon
being greater or smaller than $\rc$. $\C$ is positive for $r_+ <\rc$
and negative otherwise.

\bs
%
%
%

\bs

\no {\bf (3)} Let us now study asymptotically anti-de-Sitter \sc \bh in
arbitrary dimensions.
The critical value we obtain for the \sc horizon is
\be	\rc = {l\over{\sqrt{D-1}}}.	\label{racri}  \ee
 From relation (\ref{constraint}) the critical mass is
\be	M^c = {l\over 2} {D^{(D-2)/2}\over{(D-1)^{(D-1)/2}}},	\ee
and Eq.(\ref{tempq0}) gives us
\be  T^c_{Q=0} = {1\over{\pi(D-2)\rc}} = {\sqrt{D-1}\over{\pi (D-2)l}}
= {1\over{2\pi M^c (D-2)}}({D\over{D-1}})^{D-2\over 2}. \label{tecri} \ee
The critical entropy at the horizon is given by
\be	S^c (r_+) = \pi l^2 [ ({D\over{D-1}})^{(D-2)/2} - 1].	\ee
The critical entropy is a monotonically increasing function of
$D$, which reaches its maximum value when the dimensionality approaches
infinity
\[	S^c (r_+) = \pi l^2 [\sqrt{e} -1].  		\]

The sign of the thermal capacity is determined by the sign of $\D$
\be
\D={(D-1)\over l^2}r_+^{D-2}[r_+^2-{l^2\over{D-1}}](1+
{r_+^2\over l^2})^{(D-2)/2}.
\ee
We see that there is only one transition, which takes place
at $\rc$ given by (\ref{racri}), and the heat capacity goes from
negative to positive values as $r_+$ increases.
As in the asymptotically \RN flat case, $C_Q$ is infinite and the
rest of the thermodynamical variables remain finite during the transition.
Thus we can characterize it as a second order phase transition.

\bs
Following Hawking and Page \cite{hawpa} we use the Helmholtz free
energy to characterize the equilibrium of the system. In anti-de-Sitter space,
the gravitational potential causes confinement of particles and one can
consider the system formed by the black hole and radiation as confined
in a box. In addition, we
are taking the superradiant effects to be negligible, and the \sc to
be in equilibrium with the thermal bath.
The free energy can be written as
\be
F= M-TS = {l^2\over 2(D-2)r_+}[1+(D-1){r_+^2\over l^2}-(1+
{r_+^2\over l^2})^{D/2}], \label{helmholtz}
\ee
where we have used the expressions of $M$, $T$ and $S$ coming from
(\ref{constraint}), (\ref{tempq0}) and (\ref{entropy}) respectively.

The expression for the $D=4$ free energy is
\be
F_{D=4}={r_+\over 4}(1-\ram),
\ee
with $r_+^{F=0} = l$ the only strictly positive root.

For the particular case of four dimensional spacetime, we find
the following critical results: $\rc = l/\sqrt 3$, $M^c= 2l/3\sqrt 3$,
$T^c= \sqrt 3/(2\pi l)=1/(3\pi M^c)$ and $S^c (r_+) = \pi l^2/3$.
Since $\L<0$ we can write $l=\sqrt{a/|\L|}$, $\rc =\sqrt{a/3|\L|}$,
$T^c= \sqrt{3|\L|/4\pi^2 a}$ and $S^c (r_+) = a\pi/3|\L|$,
which agrees with the results obtained in \cite{hawpa}, provided we take
$a=3$ and the mass and entropy of the BTZ
black hole in units of $m_p^2$, where $m_p$ is the Planck mass.

Some interesting results in $D$ dimensions can now be obtained.
First of all, there is a minimum temperature
a \sc anti-de-Sitter black hole can have in any dimension. This can
easily be seen from Eq.(\ref{tempq0}) upon derivation. The minimum value
of the temperature turns out to be at $\rc$, and thus is given by
relation (\ref{tecri}).
Above $T^c$, there are two possible black hole radii
for each temperature, respectively smaller and larger than $\rc$. The
black hole with smaller horizon lies in the region $C_Q <0$ and therefore
is unstable decaying into pure thermal radiation or to a larger black hole
state. The larger black hole has $C_Q>0$ and hence is locally stable, although
we need to study the free energy to determine its behavior.

The roots and sign of Eq.(\ref{helmholtz}) are determined by
\be
1+(D-1){r_+^2\over l^2}=(1+{r_+^2\over l^2})^{D/2}.	\label{nose}
\ee
We can show that the above equation has only one root $r_+^{F=0}$
because both sides are monotonically increasing functions of $r_+$,
and the LHS is greater than the RHS for small $r_+$, and smaller for
large $r_+$.
Then, $F$ is positive for $r_+<r_+^{F=0}$ and negative for
$r_+>r_+^{F=0}$.
By evaluating the free energy at $\rc$ we obtain $F(\rc)>0$ for all $D$,
and therefore $\rc<r_+^{F=0}$.
For general $D$, if $T(\rc)\leq T \leq T(r_+^{F=0})$ the heat
capacity and free energy
are positive. For a given temperature the largest black hole solution
tends to reduce
its free energy by evaporating completely. For $T>T(r_+^{F=0})$, the
free energy of the black hole will be less than that of the
pure radiation, which will then tend to tunnel to the black hole state.
Similar effects have been  previously noticed in the literature
for $D=4$ \cite{hawpa}.

\bs
\no {\bf (4)} Asymptotically anti de-Sitter \RN in $D$ dimensions.
The roots of the denominator of equation (\ref{heat}) are given by
\be
\bigl[{D-1\over l^2 }\rc^D-\rc^{D-2}\bigr](1+{\rc^2\over l^2})^{(D-2)/2}+
Q^2\bigl[{(2D-5)\over l^2 }\rc^2+D-1\bigr]=0
\ee
or, equivalently by
\be
f(\rc) = \rc^{(D-2)/2},	\label{eqf}
\ee
where
\be
f(r_+)=Q^2\bigl[D-1+{(2D-5)\over l^2 }r_+^2\bigr]+\sum_{k=1}^{2D}
{{D/2}\choose k}(2k-1)\Biggl({r_+\over l^2}\Biggr)^k \;r_+^{(D-2)/2}.
\ee
This is a polynomial with strictly positive coefficients and
there will be then two solutions for values of $Q$ such that
$0\le Q<Q_{crit}$, and one solution for $Q=Q_{crit}$,
where $Q_{crit}$ is the value of the charge for which
$f(\rc)=\rc^{(D-2)/2}$ and $f'(\rc)=(D-2)\rc^{(D-4)/2}/2$.
For $Q>Q_{crit}$ there will be no transitions.

As in previous cases, the full thermal capacity is infinite and it
corresponds to second order phase transitions.

%
%
\bs
It is possible to show for sufficiently large values
of $Q^2/M$ that a naked singularity arises.
The condition for this to occur turns out to be
$Q^2/M > (Q^2/M)_{\rm max}$, where
\be
(Q^2/M)_{\rm max} = {2(D-3)\over{D-2}} r_{+{\rm max}}^{D-3}
\frac{1+(D-1)r^2_{+{\rm max}}/l^2}{1+ 2 r^2_{+{\rm max}}/l^2}.
\ee
and $r_{+{\rm max}}$ is the radius for $(Q^2/M)_{\rm max}$,
where the inner and outter horizons are degenerate and $T=0$.
We also found, as in the asymptotically flat case, a limiting value of
zero point entropy,
so the black hole ground state is expected to be degenerate.

\begin{center}
\section*{IV. Odd dimensional black holes}
\end{center}

In odd dimensions $D=2n-1$ the action (\ref{action}) is constructed
by considering the Euler density (basically the product of $n$ Ricci tensors)
in $D+1$ dimensions. This density
$\varepsilon_{D+1}$ is an exact form and thus can be written as the exterior
derivative of a $D$-form
\be
\varepsilon_{D+1} = \k \e_{a_1\cdots a_{D+1}} R^{a_1 a_2}\we\cdots
\we R^{a_D a_{D+1}} = d{\cal L}_D,
\ee
with
\be
{\cal L}^{\rm odd}_D = \k \sum_{p=0}^{n-1} \a_p\e_{a_1\cdots a_D}R^{a_1 a_2}
\we\cdots\we R^{a_{2p-1} a_{2p}}\we e^{2p+1}\we\cdots\we e^D.
\ee

The Lovelock coefficients are those given in (\ref{coeff}).
The equations of motion are given by
\be
(R^{a_1 a_2}+ l^{-2} e^{a_1}\we e^{a_2})\we\cdots\we
(R^{a_{D-2}a_{D-1}}+ l^{-2} e^{a_{D-2}}\we e^{a_{D-1}})\we e^{a_D}
\e_{a_1\cdots a_{D+1}}=0.
\ee

\bs
As in the previous section, to study odd dimensional \bh
we start from the expression of the metric (\ref{interval}),
where
\be     g^2(r) =  1 + \rat - [ M + 1 - {Q^2\over {2 (D-3)\,
r^{D-3}}}]^{2\over{D-1}}, \;\mbox{with $D$ odd.}    \label{metric2}  \ee
The value of the horizon $r_+$ is obtained from
\be	1+\ram = [M+1-{Q^2\over{2(D-3)\,r_+^{D-3}}}]^{2\over{D-1}}.
\label{constraint2}	\ee
The relation above tells us that there are no \sc \bh in odd dimensions
when we restrict ourselves to the case of zero cosmological constant, the
reason being that $M$ becomes zero.
However, for the case of $l$ finite, there is
the possibility for such black holes, since in this case $M$
takes positive values.

As we did in Section III, we will write the general expressions
for the temperature and entropy of the black hole as a function
of the dimensionality. From (\ref{def}) and (\ref{metric2}), the
temperature is found to be
\be T = {1\over 4 \pi} \lbrack {2r_+\over l^2}-{Q^2\over(D-1) r_+^{D-2}}
(1 + \ram)^{3-D\over 2} \rbrack,        \ee
where we have used again the implicit approach coming from the constraint
equation (\ref{constraint2}). Here one can see that for $l\to\infty$
and $Q\ne 0$ we arrive at a negative value of the temperature, for
any value of $D$.

For the case of \sc black holes, the relation found in \cite{BTZ} is
recovered,
\be     T_{Q=0}= {r_+\over 4 \pi l^2}.  \label{tqz} \ee

In odd dimensions the expression for the entropy of the black hole
given in \cite{BTZ} does not have a close form, being
\be     S= 2 \pi (D-1) \int_0^{r_+} ds (1+ {s^2\over{l^2}})^{D-3\over 2}.
\ee
For the case of $D=4$ the previous expression reduces to the
usual $S= 4\pi r_+$.

\bs
\subsection*{Phase transitions}

\bs
The full thermal capacity at constant $Q$ is given by
\be     \C = {T\over{\hat\D}}[8 \pi^2 (D-1)(1+\ram)^{D-2}r_+^{D-1}],
\label{calo}    \ee
where
\be
\hat\D = {2\over l^2}\; (1 + \ram)^{D-1\over 2}\; r_+^{D-1}
+ {Q^2\over{D-1}}\;[{{2D-5}\over{l^2}}{r_+^2} + D-2].
\ee
We are interested in the following cases:

\bs

\no {\bf (1)} Asymptotically anti de-Sitter \sc black holes. $\hat\D$
reduces to a polynomial expression in $T$, which is finite and non-zero
for all $T\ne 0$. This implies that
$\C$ is finite and regular for any temperature. Therefore there are no
critical transitions.

\bs

\no {\bf (2)} The general case, $l$ finite and $Q\ne 0$, does not present
any transitions either. It is easy to see, by inspection of (\ref{calo}),
that there is no transition with $\C$ divergent ($\hat\D\ne 0$) for any
value of $T$, $Q$ or $l$. For finite values of the heat capacity
there is no transition either, as the derivatives will be regular to any
order. This can be seen from the fact that one
can write
\be {{\pa \C}\over{\pa T}} = {{\pa S}\over{\pa T}} + T {\pa\over{\pa T}}
({{\pa S} \over {\pa T}}). \label{derivative}       \ee
The first term on the right-hand side is equal to $\C/T$ and it is
finite, since it is the ratio of two
polynomials in $r_+$ with positive coefficients. The second term
reduces to
\[	{1\over\hat\D} {\pa\over{\pa r_+}}({\C\over T}),	\]
where the ratio $\C/T$ and its derivative with respect
to $r_+$ are both finite. Because $\hat\D$ never vanishes in odd
dimensions, the second term in (\ref{derivative})
is also finite. Doing a similar analysis for the
following derivatives of $\C$ we find them all to be finite.
Therefore in  odd-dimensionally continued BTZ \bh there
are no critical transitions; the full thermal capacity and all
its derivatives remain finite for all values of $T$ and $Q$. This result
is general for any dimension.

\begin{center}
\section*{V. Conclusions}
\end{center}

In this paper we studied the possibility of critical transitions
in Ba\~nados, Teitelboim and Zanelli dimensionally continued
black holes. In even dimensions there exist transitions depending
on the value of the charge and the cosmological constant.
Asymptotically flat \sc black holes do not present phase transitions
with the specific heat being always negative.
In this case $T=0$ is asymptotically reachable as the horizon approches
infinity. For the \sc anti de-Sitter
case there is one critical transition, with $C_Q$ being negative (unstable)
or positive depending on the horizon being respectively  smaller or larger
than the critical horizon. These objects present a minimum temperature,
which is different than zero.

For \RN \bh with zero cosmological constant
there is only one second order transition, while the anti de-Sitter case
presents two. A characteristic feature of the latter case is that
there could exist up to three black hole radii for a given temperature.

Odd dimensional BTZ \bh do not present critical behavior.
The full thermal capacity remains finite with finite derivatives for
any value of the temperature.

\bs
Further study is still needed to fully understand the physical
meaning of these different transitions and the scaling behavior
associated with them \cite{mp}.

\bs
\subsection*{Acknowledgments}

We would like to thank M.\ Ba\~ nados, C.\ O.\ Lousto, C.\ Teitelboim
and J. Zanelli for helpful comments and suggestions.

\vfill\eject
\ed

--

--------------------------------------------------------------
Javier P. Muniain                     |   Phone (909) 787-7356
Department of Physics
University of California, Riverside
--------------------------------------------------------------